\documentclass[multphys,vecarrow]{svmult}
\usepackage{epstopdf}
\usepackage{graphicx}
\usepackage{amssymb}
\usepackage{amsmath}
\usepackage{float}
\usepackage{subeqnar}
\usepackage{epsfig}
\usepackage{citesort}

\begin{document}
\frontmatter
\mainmatter

\bibliographystyle{unsrt}

\title*{Prediction and Mitigation of Crush Conditions in Emergency Evacuations}
\titlerunning{}

\author{Peter J. Harding\inst{1},
	Martyn Amos\inst{1,2},
	and Steve Gwynne\inst{3}}

\institute{
Manchester Metropolitan University\\
Manchester, UK\\
\and
E-mail: {\tt m.amos@mmu.ac.uk}\\
\and
Hughes Associates, Inc\\
Boulder, Colorado, USA\\
E-mail: {\tt sgwynne@haifire.com}\\
}

\authorrunning{P. J. Harding, M. Amos, and S. Gwynne}

\date{\today}

\maketitle

\begin{abstract}
	Several simulation environments exist for the simulation of large-scale evacuations of buildings, ships, or other enclosed spaces.  These offer sophisticated tools for the study of human behaviour, the recreation of environmental factors such as fire or smoke, and the inclusion of architectural or structural features, such as elevators, pillars and exits.  Although such simulation environments can provide insights into crowd behaviour, they lack the ability to examine potentially dangerous forces building up \textit{within} a crowd.  These are commonly referred to as \textit{crush conditions}, and are a common cause of death in emergency evacuations.

In this paper, we describe a methodology for the prediction and mitigation of crush conditions. The paper is organised as follows.  We first establish the need for such a model, defining the main factors that lead to crush conditions, and describing several exemplar case studies. We then examine current methods for studying crush, and describe their limitations. From this, we develop a three-stage hybrid approach, using a combination of techniques. We conclude with a brief discussion of the potential benefits of our approach.

\noindent
\end{abstract}

\section{Introduction}\label{intro}

The events of 9/11 were widely seen and examined in the safety community and beyond. The catastrophic outcome and the minutiae of the evacuation have been examined by numerous official agencies, research organizations, media outlets, as well as Hollywood. Given this, the events of the day are incredibly well known; possibly more so than any other recent event.

Tall buildings are designed based on the assumption that an evacuation is managed, i.e.  that the evacuation will take place in stages, if required, with only certain sections of the population evacuating at any one time.   The evacuation will usually take place from those floors closest to the incident, then occur from more distant floors.  This assumption is key to the successful evacuation of these tall structures; the stair capacity is calculated based on the assumption that the majority of the population follow the evacuation procedure.  This means that the stair capacity within the structure will not be sufficient for the simultaneous evacuation of the entire population.

After 9/11 the assumption that tall buildings can be evacuated in a phased and controlled manner is being questioned.  Instead, it is often suggested that evacuees will be reluctant to remain behind in a structure, fearful of a failure in structural integrity similar to that experienced in the twin towers.  Given the nature of the incident on 9/11 and the possible consequences of remaining within the building (either by choice or through compulsion), it is now suggested that residents may choose to ignore the instructions of a staged procedure and instead move to the stairwells.  This may then overload the available staircase capacity.

Given this is the case, the consequences of failure should be examined.  If there is a failure in the acceptance of procedure then either the failure should be made as graceful as possible, or measures should be taken to resolve the issue; in either case, an understanding of the consequences of failure is vital.

It should be noted that during these scenarios it is not assumed that the conditions are dependent upon the existence of panic, which is difficult to predict and rarely the dominant evacuee behaviour\cite{sime1994a}.  In reality, it has been found that panic and irrational behaviour are a direct effect of the deteriorating conditions, rather than the cause of the deterioration itself.  Here we are assuming that crush conditions may develop simply because of the overloading of a route and may therefore be influenced by architectural, procedural, or behavioural factors.

One of the consequences of a full evacuation from a tall structure, that was originally designed for phased evacuation, is the overloading of an escape route in a relatively short period of time.  One of the most dangerous consequences of such an incident is that the exits, such as those at the base of stairways, would become overloaded, leading to many evacuees arriving at a bottleneck; i.e. the exit component is used above and beyond its design capacity.  This may then lead to conditions similar to those observed at the Rhode Island\cite{stationnightclub} and Gothenburg\cite{gothenburg} incidents, where crush incidents and falls were evident and lead to blocked egress routes and injuries.  It is therefore critical for the safety of tall structures to develop an understanding of: (1) Exactly when these conditions may develop?  (2) What factors need to be present in order for crush conditions to occur?  (3) When do these conditions become critical?  (4) How can we establish the possible consequences of this type of incident and design against them?

Here, we outline a program of work that will enable the assessment of architectural and procedural designs in order to establish whether they are prone to crush conditions developing in certain scenarios, what the consequences of this might be, and how we might best mitigate against this event.  The development of a similar tool is mentioned in the recommendations within the 9/11 report\cite{ncstar1}:

\begin{quote}
NIST recommends that tall buildings be designed to accommodate timely full building evacuation of occupants when required in building-specific or large-scale emergencies such as widespread power outages, major earthquakes, tornadoes, hurricanes without sufficient advanced warning, fires, explosions, and terrorist attacks.  Building size, population, function, and iconic status should be taken into account in designing the egress system.  Stairwell capacity and stair discharge door width should be adequate to accommodate counter-flow due to emergency access by responders.
\end{quote}

Improved egress analysis models, design methodologies, and supporting data should be developed to achieve target evacuation performance for the building population by considering the building and egress system designs and human factors such as occupant size, mobility status, stairwell tenability conditions, visibility, and congestion.

Although numerous egress models exist that are able to simulate general movement, none are able to simulate all of the conditions highlighted in NIST recommendations, along with a comprehensive crush model.  Developing such a model, that is publicly available and that can be embedded into existing egress tools, meets an identified need and will allow for a broad and vital examination of these situations.

\section{Definition of Crush Conditions}

There are many factors that play a part in the initial formation of crush conditions during an evacuation, these can be classified under the broad headings of spatial, temporal, perceptual, procedural, and cognitive components.

	\subsection{Spatial}\label{spatial}

	The spatial components of crush conditions are the simplest to quantify.  They relate to the ratio of space available for egress to the number of persons that are expected to use the escape routes.  Fruin defined this metric as the ``level of service''\cite{fruin1971a}, and highlighted the level at which the population density has the potential to facilitate the formation of crush as ``Level of Service F'', which is the density at which a single individual would have, on average, less than 0.46$m^{2}$ of space available to them.  It should also be noted that the International Maritime Organisation ({IMO}) considers an evacuation to be unsafe if, for 10\% of the overall evacuation time, the density of the evacuees reaches 4 persons per square metre\cite{IMO1033}.  This is due to the fact that, even at relatively low levels of force, prolonged exposure to ``light'' crush conditions may still cause serious injury or death.

	\subsection{Temporal}

	Temporal factors of egress vary, and depend heavily upon the rate at which conditions change.  The {RSET} (Required Safe Egress Time), defined as the elapsed time between the initialisation of an evacuation and the final evacuee reaching safety\cite{simeasetrset}, i.e. the time required for a complete evacuation under ideal circumstances.  The {ASET} (Available Safe Egress Time), defined as the total time \textit{available} for evacuation\cite{simeasetrset}, is a far more specific metric, as it will vary depending on the catalyst for evacuation (i.e. the nature of the emergency).  Traditionally, the RSET and ASET metrics have been used to determine whether or not the occupants of a building could evacuate under specific conditions.  Generally, a structure could be considered safe if the ASET value exceeds that of the RSET, i.e. there is more time available for an evacuation than would be required.  The rate at which conditions change can compound time constraints, as the ASET calculation will change dynamically with the changing conditions.  The Rhode Island nightclub fire (see Section~\ref{RI}), is a good example of this, and shows how the rapidity with which an incident escalates can place severe time constraints on the evacuating population.

	\subsection{Perceptual and Cognitive Factors}

	Perceptual and cognitive factors that lead to the formation of crush conditions are intrinsically linked, as an individual must rely on their perception of events in order to decide upon a course of action.  The individuals' perceived level of threat plays a large part in this, as it has the most direct effect on the decision making process.  Whilst the perception of threat plays a great part in the decision making process, the relationship between these two factors is highly complex, and can result in individuals displaying a wide range of behaviour, from the altruistic at one end of the scale, right through to highly competitive egress behaviour, e.g. running, pushing, etc.
	
	The perception of information also plays a key part in the formation of crush.  During emergency situations, it is often found that information relating to the current conditions is slow to propagate throughout a crowd of people, e.g. the evacuees that are placed further back in a crowd may not be aware of the conditions further ahead.  This has been found in many situations, such as the Hillsborough disaster (see Section~\ref{hills}), where the people attempting to enter a structure were unaware of the already dangerously overcrowded conditions that existed inside.  In these cases the persons at the rear of a crowd can compound the situation by producing additional force that will propagate forward through the crowd, and also by limiting the extent to which the pressure could be alleviated by inadvertently blocking the most immediate exit routes.

	\subsection{Procedural}
	
	The procedural components of crush were already alluded to (see Section~\ref{intro}), and centre around the inability, or unwillingness, of evacuees to follow strict evacuation plans in emergency situations.  This type of problem is extremely common in public buildings, where a great number of the occupants will be unfamiliar with the structure and have little, or no, knowledge of the evacuation plans, e.g. hospitals, town halls, museums, stadiums, etc.  When an evacuation takes place under these circumstances the crowd will often attempt to leave by the most familiar route, generally the route by which they entered, even though there may be exits in closer proximity.  An example of this type of behaviour can be found in the Rhode Island nightclub incident (see Section~\ref{RI}), where the majority of the crowd converged at just one point of escape, even though there were numerous other exits available.

	\subsection{Summary}
	
	The formation of crush conditions within crowds is a highly complex, emergent phenomena, and the causes of this cannot be explained by simply attributing it to the presence of panic within the crowd, which is widely regarded as being somewhat of a fallacy.  We suggest that crush conditions can only be reliably defined as a combination of all the factors mentioned above, which culminate in the individuals' inability to fully control their direction and speed of movement, thus leading to an increase in the physical forces that they are subject to.

\section{Case Studies}

Here we present case studies representing situations where the formation of crush conditions have led to both serious injuries and fatalities.  Each case study also represents some failure within a system (e.g. failure to limit the capacity of a structure to safe levels, failure to adhere to official guidelines or fire laws, failure to follow crowd control policies, etc).  These types of failure are often observed in cases where the evacuation of a building leads to the death or injury of many people. Failures of this kind are common, and we believe that they should be expected, and be considered during the design of buildings, the creation of evacuation plans, and especially in simulated evacuation exercises.

	\subsection{Rhode Island Nightclub}\label{RI}
	
	The Station Nightclub, Rhode Island, was the scene of a tragedy when, on February 20th 2003, a fire during a rock concert caused 100 fatalities and significant injuries\cite{stationnightclub}.  The fire started when the band's pyrotechnics ignited the flammable soundproofing foam that surrounded the stage, and quickly filled the club with dense, choking smoke.  The fire spread from the stage, igniting a  large portion of the ceiling, and within five minutes of the initial ignition those outside the club observed flames breaking through a portion of the roof.
	
	Despite the existence of four possible exits, the majority of the crowd headed for the most familiar exit; the entrance to the club.  This exit point was soon overwhelmed, and people began to trip or fall during their escape.  The official time-line of the fire (compiled by NIST\cite{stationnightclub}) states that just 1 minute and 42 seconds after the start of the fire, there existed a ``pile'' of people, blocking the main exit and making further egress through that route impossible.

	\subsection{Gothenburg Dancehall}
	
When fire broke out in a dancehall in Gothenburg, Sweden, on October 28th 1998, it claimed the lives of 63 people and injured over 180 others.  The first floor venue in question was packed to over double its 150 capacity, with officials estimating that there may have been over 400 people in attendance.  Eye-witness accounts of the incident state that population density prior to the start of the fire was already at dangerously high levels, with a number of the occupants observing that there were so many people present that they were unable to dance\cite{gothenburg}.  Shortly before midnight, a fire was discovered in one of the two stairways leading out of the first floor dancehall, and those near to the affected area began to evacuate.  No announcement was made to the remaining occupants, and some survivors who had been at the far end of the hall when the fire was initially discovered stated that they smelled smoke but had initially believed it to be cigarette smoke and felt no need to evacuate.  As the full evacuation began, the one remaining exit to the building quickly became overwhelmed, and the mass of evacuees began to trip or fall over others, further diminishing the capacity of the exit.

	\subsection{E2 Nightclub Incident}
	
	In Chicago's E2 nightclub on Feb 17th 2003, the security guards' use of pepper spray, to intervene during an altercation, became the catalyst for an evacuation that claimed the lives of 21 patrons\cite{epitomeNYT}.  When the security guards released the pepper spray in an enclosed space, the effects of the chemical compound on the surrounding crowd were significant.  Those close to the attack began to rush toward the exit in an attempt to escape the pepper spray, which by this point was already spreading around the club.  As the initial wave of evacuees made their way through the club, those who had not witnessed the incident began to fear for their safety, especially as it became obvious that some form of chemical agent was present.  

	Within seconds the entire crowd, consisting of over 1500 people, rushed towards the main exit.  The door to the street opened inwards, whilst the door leading to the dance floor opened outwards.  As people rushed from the club, the upper door flew outwards, pushing those on the upper landing down the steep flight of stairs.  As more people attempted to exit, they were forced on top of the fallen evacuees, and the bodies began to ``stack up'' and block the exit.  It was the tremendous pressure placed upon the fallen evacuees that caused the 21 deaths during this incident.  The most common cause of death was asphyxiation.

	\subsection{Hillsborough}\label{hills}

The Hillsborough disaster\cite{hillsborough} (Sheffield, UK), claimed the lives of 96 people and caused the hospitalisation of a further 300.  Due to the heightened public interest in the incident (the match had been transmitted live on English television), and also because of the multiple perceived failures on the part of the authorities, the Hillsborough disaster has become one of the most thoroughly investigated crowd disasters in living memory.

The tragedy at Hillsborough stadium occurred when police stewarding the match made the decision to open an extra set of gates, intended as an exit, in order to relieve the extreme levels of congestion that were forming as the crowds tried to enter the stadium through the turnstiles at the Lepping's Lane end of the ground.  These gates did not have turnstiles, and the result was an influx of up to 5,000 fans through the narrow corridor that lead into the standing terrace.  The sudden arrival of so many additional fans pushed the capacity of the central pens far above their legal maximum, and soon a dangerous crush formed at the front of the stands.  Those fans still entering the stadium were unaware of this, and continued to attempt to enter the stand as the people inside were slowly crushed against the crowd barriers and fences at the front of the stands.  The conditions at the front of the terrace became so bad that most of the 96 victims died from asphyxiation, or other crush related injuries, within five minutes of the game starting.

\section{Previous Work in the Field}

In general, each crush detection method that has been used to date can be classified into one of two generic groups; explicit methods and implicit methods.   These two generic methodologies are outlined below, along with a brief discussion of their relative strengths and weaknesses.

	\subsection{Implicit}

	The implicit methodology is the original crush detection approach, and is still highly popular, being used in a large number of simulation models\cite{erica28egress}.  This methodology relies on the expert analysis of factors such as population density (see Section~\ref{spatial}), behavioural analysis, and environmental considerations.  The analysis of conditions within these models, therefore, is left to the engineer, who interprets the output of the simulation to determine whether crush conditions have occurred.

	Implicit modelling does not take into account the possibility that evacuees may exhibit any competitive egress behaviours (e.g. pushing), as there is no accurate method for simulating these behaviours without the inclusion of force calculations.  This makes it ideally suited for general evacuation simulations; i.e. timely evacuations under ``ideal'' conditions.
	
	As the exact force being exerted upon individuals is never calculated, the precise physical danger that may exist in the evacuation can never be quantified.  The only assertion that can be made, based on an implicit analysis, is that crush conditions \textit{may} form during the evacuation in question.  The benefit of this approach is that, as the physical force calculation are not performed, it requires far less processing power than other methods.  

	There are too many implementations of the implicit methodology to list here but a popular, well documented example is Simulex\cite{simulex}, from Crowd Dynamics Ltd.

	\subsection{Explicit}\label{explicit}
	
	The explicit modelling of crush conditions incorporates an assessment of crush into the model itself, and therefore requires less user analysis than the implicit approach.  Often based on the calculation of Newtonian force values, and generally operating in 2-dimensional space, explicit methodologies may be used to detect the presence of crush conditions much more precisely than would be possible with implicit modelling techniques.  By simulating the exact forces being exerted by each individual, and enabling the propagation of forces throughout a crowd, the explicit methodology can be used to measure the exact amount of force that any individual is subject to.  This, therefore, offers the possibility of \textit{quantifying} the dangers that individuals may face, which is not possible using the implicit modelling techniques.

	Whilst the explicit methodologies offer an accurate measure of the forces acting within a crowd, the calculations needed to measure force require much more processing power than an implicit implementation, so there exists a definite trade-off between the two techniques.

	The most well-known implementation of this methodology is the Social Forces Model\cite{helbing2000}, which combines the force equations mentioned above with the modelling of the social forces acting within crowds.  Although the original Social Forces Model was created as a learning tool, rather than an full-featured simulation environment, the model has recently been incorporated into the {{FDS}+Evac} Simulation environment\cite{fdshelbinginc}.

\section{Our Proposed Approach}

We propose a three stage approach to this problem, consisting of separate processes for the \textbf{identification}, \textbf{qualification}, and \textbf{quantification} of crush conditions.  By employing different methods for all three stages of the analysis, we believe that the entire process may be completed at relatively low computational expense.  We hope to implement these techniques as part of a suite of applications, that would offer existing egress simulations the possibility of including either full or partial crush analyses, depending on the level of accuracy required.

Two of the three techniques that we propose are still relatively novel and untested, so will require validation before they would be suitable for integration into existing environments.  Each methodology will be fully tested as stand-alone applications, but a full validation will be required before the concepts are proven.  At present, the team intends to attempt to integrate the applications into the open source simulation environment {{FDS}+Evac}, to enable full validation of the models, including historical data validation and peer validation\cite{validation}.

	\subsection{Identification}
	
	In order to first identify crush conditions, we propose treating their formation as a simple phase transition, similar to those found in many social and biological systems\cite{antphase}.   In many of these systems a point is reached at which a change (often an abrupt change) can be observed, this change is characterised as a movement away from one general rule of system behaviour to another, different set of observable behaviours that dictate the state of the system as a whole.
	
\begin{figure}[h!]
	\centering
		\includegraphics[width=0.8\linewidth]{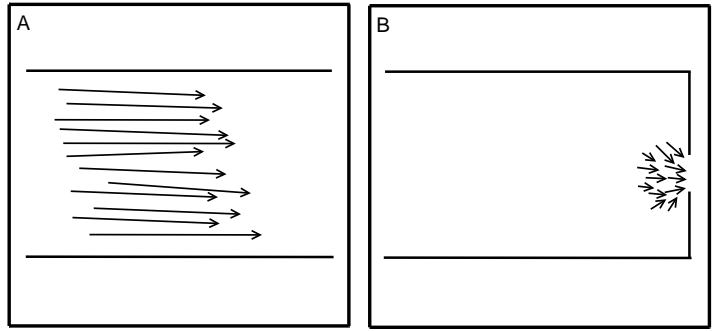}
	\caption{Slide A shows an example of the movement vectors of evacuees during the ordered state of the system, with all vectors showing a good deal of similarity.  Slide B shows example vectors during the disordered state, with the vectors varying a great deal more in both direction and magnitude}
	\label{fig:vec}
\end{figure}

	In egress situations, a crowd will usually head towards the most familiar exit, often forming groups either before or during this action.  The evacuees that make up these groups will have similar trajectories to their closest neighbours and will be travelling at a similar speed (i.e the flow, within each group, can be considered laminar).  This would form the general rule for the ordered state of this system (see Fig~\ref{fig:vec} - A).  If the evacuees are impeded in any way during their exit (e.g. they come across an obstacle in their path, or reach a congested area), they will reduce their speed and be forced to change their directions of movement, or forced to remain stationary (i.e.  the flow becomes non-laminar, or turbulent).  This would form the general rule for the disordered state of this system (see Fig~\ref{fig:vec} - B).

	Buckingham's $\Pi$ Theorem\cite{buckingham} is a key theorem in dimensional analysis, and can be used to create a set of dimensionless variables that allow the analysis of an unfamiliar system, i.e. a system for which the equations governing its behaviour are either partially or wholly unknown.  We will apply this theorem to the agent data within an egress model, to reduce the system to a number of dimensionless quantities, which can then be analysed to ascertain the state of the system at any one time.  The advantage of this approach is that both the agent's physical variables (e.g. speed, direction, mass) and their decision making variables (e.g. perceived level of threat, tendency toward competition) are considered, which will provide a more comprehensive analysis of crush than could be achieved by movement variables alone.
	
	Further analysis is achieved by the use of Mutual Information (MI)\cite{mi}, a technique that has been used to quantify the similarity of two signals.  This methodology was first used by Wicks \textit{et al}\cite{wicks} to detect phase transitions within a well-known flocking model\cite{vicsek}, and was found to accurately identify the point of phase transition even when only a subset of the agents' data were analysed.  We will employ a similar methodology to detect the formation of crush conditions within localised groups of agents, using the MI method to quantify the extent to which our ``idealised'' (ordered) agent-state (see fig~\ref{fig:vec}), differs to that of the current state.  We will dynamically restructure agents into groups, based on their current locale, and treat each group as a system within its own right, tracking a subset of each ``sub-system'' to identify the earliest stages of crush formation without the need to track \textit{every} agent throughout the \textit{entire} evacuation.

	\subsection{Qualification}
	
	To qualify the presence of crush conditions within the crowd, we intend to use a time-series, neural network classifier\cite{genann} to analyse the agent variables and movement patterns.  This will give an indication of the amount of pressure that is likely being exerted on the individual in the crowd.  The classifier acts as a statistical data analysis tool, and is configured to recognise the conditional similarities shared by individuals affected by the onset of crush conditions.
	
	The neural network approach has been selected for two main reasons.  Firstly, after the initial training program, the neural network approach requires far less computational power to make its classification than other statistical analysis techniques, reducing the classification during normal running conditions to little more than matrix arithmetic.  The reduction in computation, in relation to other techniques, will free up system resources for utilisation by other tasks.  Secondly,  the method of classification used in a neural network is highly robust, as it does not rely on any ``system specific'' variables, which makes the deployment of this technique possible across a wide range of existing egress simulations, without the need for extensive configuration.

	By employing a time-series, neural network\cite{ANN} (i.e. a neural network that accepts input in the form of sequential data representing changes over time), we also hope to identify the qualitative similarities of individuals exhibiting competitive egress behaviour.  It will enable us to analyse growing behavioural trends, rather than just classify an agent's behaviour at one precise moment in time.

	To train the network, we will collect time-series agent data from a ``full-force'' simulation, i.e. a simulation in which a physical force model is running, which should enable the network to recognise the qualitative similarities that individuals affected by crush share.  We hope that training the network using this type of data will allow the network to associate the existence of a variety of conditions to the presence of crush, therefore negating the need to engage a physics engine for all subsequent simulation runs.

	\subsection{Quantification}
	
	To fully quantify the effects of force propagating through a crowd, a physical force model is employed, based on the explicit crush detection method mentioned previously(see Section~\ref{explicit}).  We currently plan to implement this physical force model as a rigid body dynamics engine\cite{rigbod}, with representations of such variables as mass, velocity, friction, and force propagation, modelled according to the laws of Newtonian mechanics.  The engine will solve simplified physical equations in two dimensional space, resulting in good approximations\cite{dynamics} of force calculations that can be completed in as little time as possible.
	
	The possibility of modelling this phenomena as a soft body dynamical system will be investigated, as recent research has highlighted the need to incorporate calculations for the compression forces acting within crowds\cite{chunxia2007a}, but our initial research into the feasibility of this approach leads us to believe that the calculations involved would be prohibitively computationally expensive at this time.

	\subsection{Hybrid Approach}

The methodologies outlined above may each be employed individually, to add differing degrees of crush analysis to a simulation, but we also propose a conceptual framework, within which all three methodologies could be combined to create an analytical tool that applies crush calculations intelligently.  This approach will allow us to retain the accuracy of force calculations whilst reducing the computational expense associated with it.

\begin{figure}[h!]
	\centering
		\includegraphics[width=0.8\linewidth]{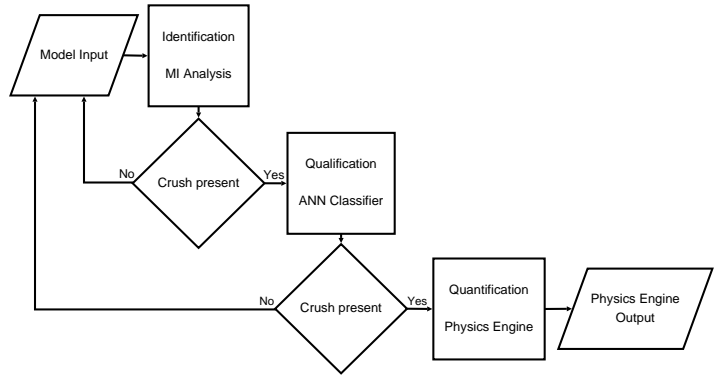}
	\caption{Process flow diagram depicting the interactions between the three application, according to the suggested framework.}
	\label{fig:iphys}
\end{figure}

The proposed approach requires the analysis of conditions based on locale, i.e. analysing conditions in different locations as if they were separate systems, and the escalation of analytical accuracy upon confirmation of crush.  Figure~\ref{fig:iphys} shows the flow of control across the three applications.

By applying the more accurate analyses \textit{only} once crush has been confirmed by the previous method, the most computationally expensive techniques will only be applied to affected areas, rather than across the entire behaviour space.  This leaves us with the possibility of having different analyses being applied simultaneously, within the same simulation, but in different geographical locales, e.g. the identification method is running on a corridor where the flow of pedestrians is laminar, whilst at the exit of a stairwell, where a crowd has formed, the analysis would be carried out by the quantification method.  The advantage of engaging each application in this way is that it will ensure that the most serious effects of crush, the build up of forces within a crowd, are measured precisely, without calculating force for all agents within the simulation.

\section{Benefits of our Approach}

This approach to crush analysis will provide a new tool, suitable for integration into existing simulation environments, that will allow engineers the ability to incorporate different levels of analysis for each specific simulation.  The inclusion of such analytical methods will add a further dimension to traditional models, and further the realism of current simulation tools.

The addition of crush analysis techniques into models will allow engineers to better test the robustness of evacuation procedures, carry out more realistic recreations of historical incidents, and more comprehensively investigate the safety of architectural designs.  It is the aim of this project to supply further tools to the evacuation sciences community that will allow this to happen, and act as a further weapon in the armoury of the engineers, technicians, and analysts that operate in this field.

\section{Conclusion}

The need for further crush analysis techniques has been clearly stated, and the phenomena that we wish to simulate precisely defined.  We have presented three methodologies for the detection, confirmation, and measurement of crush conditions within a simulation environment, and a theoretical framework within which they could operate in unison, reducing computational expense without a reduction in accuracy.

The short-term goal of this research is simply to prove the suitability of these concepts for use in the analysis of crush, by the creation of a prototype implementation that may be used for experimentation.  In the long-term we are looking to integrate this prototype into a larger simulation environment, to prove its feasibility as an ``off the shelf'' component to an evacuation model.

\section*{Acknowledgements}
	This work is partially supported by the Dalton Research Institute, Manchester Metropolitan University.

\bibliography{peter-harding}

\begin{thebibliography}{10}

\bibitem{sime1994a}
J~Sime.
\newblock Escape behaviour in fires and evacuations.
\newblock {\em Design Against Fire}, 1994.

\bibitem{stationnightclub}
W~Grosshandler, N~Bryner, D~Madrzykowski, and K~Kuntz.
\newblock Report of the technical investigation of the station nightclub fire.
\newblock Technical report, NIST, 2005.

\bibitem{gothenburg}
E~Comeau and R~F Duval.
\newblock Dance hall fire gothenburg, sweden, october 28, 1998.
\newblock Technical report, National Fire Protection Association, 2000.

\bibitem{ncstar1}
Final report on the collapse of the world trade center towers.
\newblock Technical report, NIST, 2005.

\bibitem{fruin1971a}
J~Fruin.
\newblock {\em Pedestrian Planning and Design}.
\newblock Metropolitan Association of Urban Designers and Environmental
  Planners, 1971.

\bibitem{IMO1033}
IMO.
\newblock Interim guidelines for evacuation analyses for new and existing
  passenger ships.
\newblock Technical report, International Maritime Organisation, 2002.

\bibitem{simeasetrset}
J~Sime.
\newblock An occupant responses escape time (oret) model.
\newblock In {\em Proceeding of the First International Symposium}, 1998.

\bibitem{epitomeNYT}
J~Wilgoren.
\newblock 21 die in stampede of 1,500 at chicago nightclub.
\newblock {\em New York Times}, 18.02.2003.

\bibitem{hillsborough}
The hillsborough stadium disaster: final report.
\newblock Technical report, U.K. Home Office, 1989.

\bibitem{erica28egress}
E~Kuligowski.
\newblock Review of 28 egress models.
\newblock In {\em Workshop on Building Occupant Movement During Fire
  Emergencies}, 2004.

\bibitem{simulex}
P~A Thompson and E~W Marchant.
\newblock A computer model for the evacuation of large building populations.
\newblock {\em Fire Safety Journal}, 24:131--148, 1995.

\bibitem{helbing2000}
Dirk Helbing, Illes Farkas, and Tamas Vicsek.
\newblock Simulating dynamical features of escape panic.
\newblock {\em Nature}, 407:487--490, 2000.

\bibitem{fdshelbinginc}
Juha-Matti Kuusinen.
\newblock Group behavior in {FDS}+evac evacuation simulations.
\newblock Published online, August 2007.

\bibitem{validation}
J~Kleijnen and R~Sargent.
\newblock A methodology for fitting and validating metamodels in simulation.
\newblock {\em European Journal of Operational Research}, 120:14--29, 2000.

\bibitem{antphase}
2001 | vol. 98 | no. 17 | 9703-9706 PNAS | August~14.
\newblock M beekman and d j t sumpter and f l w ratnieks.
\newblock {\em PNAS}, 98(17):9703--9706, 2001.

\bibitem{buckingham}
Malcolm Longair.
\newblock {\em . Theoretical Concepts in Physics: An alternative view of
  theoretical reasoning in physics}.
\newblock Cambridge Univ. Press, 2 edition, 2003.

\bibitem{mi}
A~Kraskov, H~Stogbauer, and P~Grassberger.
\newblock Estimating mutual information.
\newblock {\em Phys. Rev. E}, 69, 2004.

\bibitem{wicks}
R~Wicks, S~Chapman, and R~Dendy.
\newblock Mutual information as a tool for identifying phase transitions in
  dynamical complex systems with limited data.
\newblock {\em Physical Review E}, 75(5), 2007.

\bibitem{vicsek}
T~Vicsek, E~Ben-Jacob A~Czirok, I~Cohen, and O~Shochet.
\newblock Novel type of phase transition in a system of self-driven particles.
\newblock {\em Phys. Rev. Lett.}, 75:1226, 1995.

\bibitem{genann}
Mohamad~H. Hassoun.
\newblock {\em Fundementals of Artificial Neural Networks}.
\newblock MIT Press, 1995.

\bibitem{ANN}
A~Kehagias and V~Petridis.
\newblock Predictive modular neural networks for time series classification.
\newblock {\em Neural Networks}, 10(1):31--49, 1997.

\bibitem{rigbod}
J~Wittenburg.
\newblock {\em Dynamics of Multibody Systems: Dynamics of Systems of Rigid
  Bodies}.
\newblock Springer-Verlag, 2007.

\bibitem{dynamics}
R~{van Zon} and J~Scho?eld.
\newblock Numerical implementation of the exact dynamics of free rigid bodies.
\newblock {\em J. Comput. Phys}, 225:145?--164, 2007.

\bibitem{chunxia2007a}
L~U Chunxia.
\newblock Analysis of compressed force in crowds.
\newblock {\em J Transpn Sys Eng \& IT}, 7(2):98--103, 2007.

\end{thebibliography}

\end{document}